\documentclass[aps,nofootinbib]{revtex4}

\usepackage{amstext,amsmath,amssymb}
\usepackage[dvips]{graphicx}
\usepackage{latexsym}
\newcommand{\Ref}[1]{(\ref{#1})}

\def\ee{{\cal E}}

\def\mm{{\cal M}}

\def\oo{{\cal O}}

\def\tt{{\cal T}}

\newcommand{\Z}{\mathbb{Z}}

\newcommand{\R}{\mathbb{R}}

\def\be{\begin{equation}}
\def\ee{\end{equation}}
\def\bes{\begin{eqnarray}}
\def\ees{\end{eqnarray}}
\def\arr{\rightarrow}

\def\la{\langle}
\def\ra{\rangle}
\def\f{\frac}

\def\what{\widehat}

\def\arr{\rightarrow}
\newcommand{\SO}{\mathrm{SO}}
\newcommand{\lalg}[1]{\mathfrak{#1}}

\begin{document}

\title{Deformed Special Relativity as an effective flat limit of quantum gravity}

\author{{\bf Florian Girelli}\footnote{fgirelli@perimeterinstitute.ca},
{\bf Etera R. Livine}\footnote{elivine@perimeterinstitute.ca}}
\affiliation{Perimeter Institute, 35 King Street North
Waterloo, Ontario Canada N2J 2W9}
\author{{\bf Daniele Oriti}\footnote{d.oriti@damtp.cam.ac.uk}}
\affiliation{Department of Applied Mathematics and Theoretical Physics,
Centre for Mathematical Sciences, University of Cambridge,
Wilberforce Road, Cambridge CB3 0WA, England, EU}

\begin{abstract}

We argue that a (slightly) curved space-time probed with a finite resolution, equivalently a finite minimal length, is effectively described by a
flat non-commutative space-time. More precisely, a small cosmological constant (so a constant curvature) leads the $\kappa$-deformed
Poincar\'e flat space-time of deformed special relativity (DSR) theories. This point of view eventually helps
understanding some puzzling features of DSR. It also explains how DSR can be considered as an effective flat (low energy) limit
of a (true) quantum gravity theory. This point of view leads us to consider a possible generalization of DSR to arbitrary
curvature in momentum space and to speculate about a possible formulation of an effective quantum gravity model in these terms. It also leads us to suggest a {\it doubly deformed special relativity} framework for describing particle kinematics in an effective low energy description of quantum gravity.
\end{abstract}

\maketitle

\section{Introduction}

A recurring theme in quantum gravity is the existence of a universal length scale set by the Planck length or
equivalently a universal mass scale, obtained by combining the Planck constant $\hbar$ and the Newton constant for
gravity $G$. Such a length scale seems at odd with special relativity and it is thus interesting to see whether one
could write a theory of {\it Deformed (or Doubly) Special Relativity}, which would take into account both a universal speed (the
speed $c$ of light or of massless particles) and a universal length scale. A first step, done by Snyder \cite{snyder}
in 1947, was to write a Lorentz invariant kinematical framework where (some of) the space(time) coordinates have a discrete
spectrum of the type $\Z\,l_P$ (see also \cite{discrete}). He found as a natural consequence of his model that the coordinates, which are now
raised to the level of operators, are non-commuting. More recently, the idea was further studied and developed (see for
example \cite{giovanni,leejoao}) and it was shown that the underlying symmetry of Deformed Special Relativity is a
quantum group -the so-called $\kappa$-deformed Poincar\'e group. The geometrical picture is a non-commutative
space-time with a curved momentum space (either de Sitter or Anti de Sitter) \cite{snyder,jerzy}. The issue now is
whether such a Deformed Special Relativity (DSR) can be derived/considered as an effective theory for quantum gravity
in some special regime. This is such a natural idea since the $\kappa$ deformation parameter is assumed to be linked to
the Planck length (or mass) and therefore depends on the Newton constant $G$. Then it would mean that we manage to
re-absorb (basic) quantum gravity effects influencing the dynamics/propagation of particles in space-time in the kinematics of the effective theory describing the coupled particle+gravity system.

\medskip

Now, in three space-time dimensions, the link between quantum gravity and 3d DSR theory is well understood. Let us
start by reminding that: \be l_P=\frac{G\hbar}{c^3}, \qquad m_P=\f{c^2}{G}, \ee so that we can expect some (quantum) gravity effect
due to the Planck mass even at the classical level, i.e. when $\hbar\arr 0$. More exactly, it was shown that the algebra of
(Dirac) observables of one particle coupled to gravity reproduces the $\kappa$-deformed Poincar\'e algebra, at the
classical level with a vanishing cosmological constant $\Lambda=0$ (see \cite{dsr3d} and references cited therein). A
similar result is also obtained in the framework of spin foam models for quantum gravity \cite{review}: the underlying symmetry of
the Ponzano-Regge model was identified to be the $\kappa$-deformed Poincar\'e group \cite{sf3d} (more precisely the
Drinfeld double of the Lorentz group) and particles were included in the theory as representation of the deformed
symmetry group.

Another approach is to include the cosmological constant $\Lambda\ne0$ in the gravity theory. Classically, the symmetry
group becomes the de Sitter or Anti-de Sitter group, $\SO(3,1)_{\Lambda}$ or $\SO(2,2)_{\Lambda}$, depending on the
sign of $\Lambda$. When $\Lambda$ goes to 0, the group reduces to the usual Poincar\'e group. At the quantum level, it
turns out that the relevant group is a quantum deformation $\SO_q(3,1)_{\Lambda}$ or $\SO_q(2,2)_{\Lambda}$ with the
deformation parameter $q$ being related to $\Lambda$. Typically, for positive cosmological constant, $\Lambda>0$, we
have $q=\exp(-l_P\Lambda)$. So that when $\Lambda$ goes to 0, $q$ is goes to 1 and the quantum group reduces to the
classical group. Now, interestingly, when looking at the limit $\Lambda l_P\arr 0$ of $\SO_q(3,1)_{\Lambda}$ or
$\SO_q(2,2)_{\Lambda}$, we get the $\kappa$-deformed Poincar\'e group\cite{dsrcosmo}, which is then to be identified as the symmetry
group of quantum gravity in the flat regime $\Lambda=0$.

These points of view could possibly merge if we consider gravity (at $\Lambda=0$) coupled to matter fields. Looking at
the field as a made of particles, we get the first situation. On the other hand, the energy density of the matter field
would generate locally an effective cosmological constant $\Lambda_{eff}\ne 0$, which places us in the latter
framework. In the case that $\Lambda_{eff}\ll 1/l_P^2$, i.e when the back-reaction of the scalar field on the geometry
is very small, then the concept of particle would be meaningful and the two points of view would lead to the same
picture.

\medskip

In four space-time dimensions, the situation is more complex. First, the Planck mass now depends on the Planck
constant:
\be
l_P=\sqrt{\f{G\hbar}{c^3}}, \qquad m_P=\sqrt{\f{\hbar c}{G}},
\ee
so that it is not likely that DSR would arise at the classical level, but should truly be an effective description of quantum gravity effects. One could consider a regime in which the Planck length is negligible and can be set at 0 although the Planck energy is still relevant (as a maximal energy for a particle for example). More precisely, this would be a semi-classical flat limit, where both $\hbar$ and $G$ can be taken to 0, while the quantum effects and the gravitational effects are still on the same order of magnitude $\hbar\sim G$, so that $l_P\arr 0$ while $m_P$ is fixed. In such a limit, we might expect to recover the DSR framework. Now, are there more precise motivations for considering DSR as an effective theory for quantum gravity?

On one hand, Loop Quantum Gravity (LQG) derives a discrete spectrum for geometrical operators (such as areas and
volumes) at the kinematical level, which would motivate considering a flat theory with a universal length scale
(corresponding to the universal/minimal geometrical scale in LQG) as a semi-classical regime of the full theory.
Actually, LQG and DSR provide very similar frameworks to deal with discrete quantum geometry without breaking Lorentz
invariance \cite{discrete}.

On the other hand, it has also been conjectured that the symmetry of the quantum gravity vacuum at $\Lambda\ne0$ should
be the q-deformed de Sitter (or Anti de Sitter) group $\SO_q(4,1)$ (or $\SO_q(3,2)$). Then similarly to the 3d case, in
the limit $\Lambda l_P^2\arr 0$, this symmetry algebra would get contracted down the $\kappa$-deformed Poincar\'e
algebra\cite{dsrcosmo}. This would then be the symmetry algebra of the semi-classical flat effective theory for quantum
gravity, describing the perturbations around the classical Minkowski space-time.

\medskip

Let us point out that there exists other possible limits of quantum gravity, corresponding to observers looking  at phenomena corresponding to another regime where the different constants have a different relevance. For example, one could look at a Newtonian limit with $c\arr \infty$ while keeping $G$ finite. Then we could still retain some quantum gravity effect, by sending $\hbar$ to 0 while keeping $m_P$ finite. Such a limit should lead us to some corrections to Newtonian Dynamics and it would be interesting to check whether it fits with the proposed MOND framework \cite{MONDReview}.
At the end of the day, this regime corresponds to sending $l_P$ to 0 while keeping $m_P$ finite, which is the same limit that should give us DSR. Therefore, we expect a link between these two frameworks, DSR and MOND (we will say more on this possible link in the last section).

\medskip

In the following sections, we will present a general argument in favor of the idea that DSR would arise as an effective description of particle dynamics in a quantum gravity context (around the flat background). By {\it effective} we mean describing the dynamics of
particles and matter fields (and possibly of some geometrical degrees of freedom as well) living in a \lq\lq flat" background and taking into account some quantum
gravity effects (\lq\lq flat" meaning that it can be thought of as a semi-classical Minkowski space-time).

We start by reminding that the (Einstein) action for general relativity needs a length scale in order to give a
dimensionless scalar\footnotemark. As in a quantum theory of gravity, formulated for example via a path integral
approach, the quantum dynamics is encoded in amplitudes that are suitable functions of the action, this means that one
needs to choose a length scale to study quantum general relativity. It seems then natural to consider the length
resolution of the observer probing (the geometry of) space-time, i.e analyze which effective theory would an observer
with a finite resolution $\delta l$ use to describe physics in a quantum curved space-time provided by quantum gravity.
Moreover, in a semi-classical regime, quantum gravity is supposed to induce local fluctuations of the metric around the
flat Minkowski space-time. The most basic effect is to induce a scalar curvature, i.e a cosmological constant or
equivalently a non-zero vacuum energy density. Putting this two inputs together, the natural question becomes: how
would an observer with a finite resolution $\delta l$ describe the usual physics on a slightly curved space-time with
$\Lambda\ne 0$ as (modified) physics on a flat space-time? We will see that, because $\delta l\ne 0$, one can trade the
curvature of space-time for a curvature in momentum space, thus obtaining a non-commutative space-time by duality and
the DSR framework in the limit where $\Lambda\ll 1/(\delta l)^2$ (the curvature radius large compared $\delta l$), i.e.
when the cosmological constant, although present, can be neglected.

\footnotetext{The action of General Relativity (GR) is $S=\int d^4x\,\sqrt{- det({}^4g)}R$, which has the
dimension of a length squared; then we should make it into $S/l^2$ to make it dimensionless, so that one can consider its exponential as the quantum amplitude for each geometric configuration entering in the
path integral approach, for example. This procedure, therefore, introduces a length scale in the theory. Usually, one takes the universal scale $l$ to be the Planck length $l_P$, which in fact appears naturally when one divides the usual pre-factor $c^3/G$ in front of the action by the Planck constant, to write down the usual path integral for gravity.}

\medskip

This new perspective on DSR will then lead us to propose two sorts of generalizations of it. First to take into account a cosmological constant $\Lambda\ne 0$ leading to a {\it Doubly Deformed Special Relativity} where $\Lambda$ defines a minimal resolution on the momentum space. Then to deal with an arbitrary metric on momentum space, which would supposedly describe physics on a generic semiclassical generally curved quantum gravity state (similar to a a generic weave state of Loop Quantum Gravity).

We will also discuss the link between our point of view, in this last generalization of DSR, and the framework of
unimodular general relativity (see for example \cite{unimodular}).

\section{Deformed special relativity in a nutshell}

In this section we recall briefly the main features of deformed special relativity theories. They were introduced in
order to accommodate the notion of a invariant scale in the scheme of Special Relativity, without dropping the Relativity Principle \cite{giovanni}. This invariant scale is in
general identified with the Planck scale, because ideally linked with the physics of quantum gravity.

The basic issue about having an invariant length assumed to be a minimal length is the Lorentz contraction in
relativistic theories: the classical length of a ruler changes under boost. If one wants to keep this scale as an invariant,
it is natural to deform the action of the boosts, using a parameter $\kappa$, usually assumed to be ($\hbar$ times) the inverse of the Planck length, $\kappa=\hbar/l_P$, while
keeping the rotations unaltered. Modifying the action of
the boosts implies also modifying the action of the translations of the Poincar\'e group. At the algebraic level,
although the Lorentz algebra is not modified, we need to deform the action of the boosts on the translations (and
reciprocally). Translations are identified with the momenta and therefore one can reconstruct a deformed
space-time by duality.

We have the general commutation relations
\begin{equation}\label{lorentz}
\begin{array}{rcl}
[M_i, M_j]&=& i\epsilon_{ijk}M_k, \;[N_i, N_j]= -i\epsilon_{ijk}M_k, \; [M_i, N_j]= i\epsilon_{ijk}N_k, \\
{[}M_i, p_j{]}&=& i\epsilon_{ijk}p_k, \; [M_i, p_0]= 0,
\end{array}
\end{equation}
where we identified the translations with the momenta. The missing relations are deformed, and their most general form
is
\begin{equation}\label{deformation1}
\begin{array}{rcl}
[N_i, p_j]&=&A\delta_{ij}+B p_ip_j + C \epsilon_{ijk}p_k,   \\
{[}N_i, p_0{]}&=& D p_i,
\end{array}
\end{equation}
where $A, B, C, D$ are functions of $p_o, p_i^2, \kappa$. We would like that the deformed Poincar\'e group becomes the
usual Poincar\'e group in the continuum limit where $\kappa\rightarrow \infty$. This gives therefore some conditions on
these functions ($A, D \rightarrow 1$, $B, C\rightarrow 0$). We can moreover show that the function $C$ has to be zero
from the Jacobi identity, and also we must have the differential equation
\begin{equation}\label{deformation2}
\frac{\partial A}{\partial p_0}D+ 2\frac{\partial A}{\partial \overrightarrow{p}^2}(A+ \overrightarrow{p}^2B)-AB=1.
\end{equation}
Different solutions of this equation, with the limit conditions for $\kappa\rightarrow\infty$ give us different
deformations. It is interesting that those different solutions can be interpreted as different coordinates systems over
a de Sitter space of momenta, $\SO(4, 1)/\SO(3,1)$. Indeed one can start by considering a five dimensional Minkowski
space of momenta, and decompose the generators of the symmetry group $\SO(4,1)$ into the Lorentz part and the rest. The
four generators not in the $\lalg{so}(3,1)$ subalgebra are then identified with the space-time coordinates, acting as
translation on momentum space. The Lorentz part of SO(4,1), on the other hand, is acting in the regular way on the
Minkowski coordinates $\eta_j$,
\begin{equation}
\begin{array}{rcl}
[M_i, \eta_j]&=& i\epsilon_{ijk}\eta_k,  \; [M_i, \eta_0]=[M_i, \eta_4]=0, \\
{[}N_i, \eta_j{]}&=& \delta_{ij}\eta_0,  \; [N_i, \eta_0]= i\eta_i,  \; [N_i, \eta_4]=0.
\end{array}
\end{equation}
When restricting the $\eta$'s to the homogenous space $\SO(4, 1)/\SO(3,1)$, in a particular coordinates system, e.g.
$\eta_0= \eta_0(p_0, \overrightarrow{p}), \eta_i = p_i \,\eta(\eta_0(p_0, \overrightarrow{p})), \eta_4 = \sqrt{\kappa-
\sum_{i=0}^{3}\eta_i^2 }$,  one then recovers the previous commutation relations (\ref{lorentz}, \ref{deformation1}),
with the functions $A,B,D$  expressed in terms of $\eta_0, \eta$. This basic fact that we have de Sitter space as
momentum space will be justified in the next section, following more physical considerations.

One can now take advantage of the algebraic structure and define the space-time sector corresponding to this momentum
space. This is done through the so-called Heisenberg double. This technique uses the fact that the phase
space can be interpreted as a cross product algebra: a pair of dual Hopf algebra, acting over each other. One is
representing the algebra of momenta, whereas the other one is the space-time coordinates algebra. They are acting over each other by
translations.

One needs therefore to make the momentum algebra a Hopf algebra, and so to define the coproduct $\Delta$ by duality from
the product. It is clear that the shape of the coproduct will depend as well on the choice on the functions $A,B,D$. It
is then used to define the product on the space-time coordinates, as there are defined as dual to the momenta.
\begin{equation}
\begin{array}{rcl}
<p_{\mu}, x_{\nu}>&=&g_{\mu\nu}, \; \; \textrm{with } g_{\mu\nu}= \textrm{diag}(-1,1,1,1) \\
<p, x_1.x_2> &=& <\Delta p,x_1\otimes x_2 >.
\end{array}
\end{equation}
The result is a non commutative space, which has a different non commutativity according to the chosen momenta
coordinates system. Obviously they are all related in the same way the different coordinates systems are related. The
usually pinpointed coordinates are the $\kappa$-Minkowski coordinates $x$,
or the Snyder's coordinates $X$, satisfying:
$$
[x_0, x_i]= -\frac{i\hbar}{\kappa}x_i, \; \; [x_i, x_j]= 0,
$$
$$
{[}X_{\mu}, X_{\nu}{]}= -\frac{i \hbar^2}{\kappa^2}J_{\mu\nu},
$$
where the $J_{\mu\nu}$ are the Lorentz generators.

\section{From curved space-time to a flat non-commutative space-time}

In this section we would like to show how we can interpret a slightly curved space-time as a non-commutative space-time
and vice-versa. They are a priori different mathematical objects, however if one adds the information of a minimal
length scale, noted $\delta l$, then one can show that they lead to equivalent physics. The  argument we are giving is
some kind of extension of the argument used in field theory: a minimal resolution in the coordinate space is equivalent
to a UV cut-off in momentum space.
A minimal resolution being present, we do not have access to the differential structure anymore and the  momentum space, not required to be the flat (co-)tangent space of the manifold, might acquire a non-trivial curvature. We will make this idea more precise and show that the minimal resolution, together with the curvature of space-time, will actually generate a non-trivial curvature on the momentum space, which we will evaluate in terms of the known quantities.

Let us start with a space-time manifold ${\cal M}$ of constant curvature given by the cosmological constant $\Lambda$.
Let us take $\Lambda>0$ for simplicity; all the following arguments would obviously hold for the Anti de Sitter space
with $\Lambda<0$. One should keep in mind that in a background independent theory like General Relativity, there is no
preferred scale of reference, so that if one is working in the context of a constant curvature space then one is always
free to renormalize this curvature to any value. One needs to define some length unit with respect to which the
curvature radius will be given.

To be more precise, let us consider an observer. From the experimental point of view, it is natural to assume that this
observer has access only to a finite resolution (in length), which we note $\delta l$. From the theoretical point of
view, usual arguments in quantum gravity tell us that $\delta l \gtrsim l_P$ where the (4d) Planck length is
$l_P=\sqrt{G\hbar/c^3}$. Such a minimal resolution provides us with a natural scale and it is then natural to express
the curvature radius in this unit. For example, if  space-time is curved at a smaller scale than $\delta l$ then the
resolution will not allow to measure this curvature. We are now interested in a small curvature regime. Explicitly, the curvature radius is the cosmological length $l_c=1/\sqrt{\Lambda}$ and we are interested in the domain $l_c\gg \delta l$. We would like to formulate an effective theory corresponding to what the observer will see in such a
space-time.
More precisely, as the curvature is very small and undetectable given
the resolution available to him, the observer can naturally
mistake it for a flat space-time; however there would still be effects due
to the non zero curvature and the non zero resolution $\delta l$ and
the question is how to take them into account in an effective
description. Assuming that quantum gravity generates $\Lambda$ through
quantum
fluctuations of the geometry around the quantum Minkowski vacuum, then
such an effective theory is to be considered as the effective flat limit
of quantum gravity. We argue below it turns out to be described by a
non-commutative
(Minkowski) space-time, more precisely with a $\kappa$-deformed Poincar\'e
symmetry, and by deformed special relativity physics.


\medskip


Let us start with the curved space-time ${\cal M}$ with constant curvature given by $\Lambda$.
The observer, located at the point $x$, ignoring the curvature of space-time will think about it as the flat Minkowski space-time $M$, which can be identified to the tangent space $T_x\mm$ (up to scale factors). Of course, the equivalence principle tells us that the space-time is locally flat, meaning that the observer can consider it flat at the chosen point $x$ but will see deviations from the flat metric already around the point $x$. Here the starting point is assuming that the observer assumes the space-time flat in a (big enough) neighborhood of the point $x$: then how will he take into account the deviations from the flat metric in his description of physics? This extended flatness hypothesis is natural for an observer with finite resolution $\delta l\ne0$ since it doesn't make much sense anymore to speak about a point. Let go further and think about the motion of a particle in such a setting. When considering a minimal change of position $\delta x$ of the particle, we do not have access anymore to the infinitesimal calculus of differential geometry since the observer's resolution is limited by $\delta l$.
Therefore we can not consider $\delta x$ infinitesimally small but need
to consider it of finite size therefore as a real small
displacement on the original curved
space-time ${\cal M}$; although assuming the space-time is now the
flat space-time $M$, the observer will still consider small coordinate
variations that probe a curved space-time. Now, a flat space-time is
identified with its tangent/momentum space, i.e. it is characterized by
the
fact that it is trivially isomorphic to its tangent space (in turn
trivially identified with momentum space), as they are all modeled by
$\mathbb{R}^4$ with the Minkowski metric, so that coordinate variations
are trivially generated by momenta: $\delta x_\mu\propto p_\mu \tau$,
where $\tau$ is the (dimensionless) amplitude of the variation.
Having a curved metric on the $\delta x$'s means to have naturally the same curved metric (up to scale) on the momentum space: the observer with finite resolution works with a flat space-time but a De Sitter momentum space. Taking in account the dimension of the $x$ and $p$'s, we see that $\delta l$ and $\Lambda$ are essential for the argument since we need to insert them in the previous relation:
$$
\delta x =\delta l \f{p}{\sqrt{\Lambda}\hbar}\,\tau.
$$
Then, to extract the value of the curvature $\kappa$ of the momentum
space, we think about it as the dual of the original space-time $\mm$
under the Fourier transform. Indeed the usual argument in field theory
is that having a minimal resolution in the coordinate space implies a
UV cut-off in the momenta space: it does not make sense to talk about momenta of norm larger than $\hbar/\delta l$. Then $\kappa$, being the ``cosmological radius'' of the De Sitter momentum space, is the bound on momenta which are accessible to (or, in other words, measurable for) the observer:
\be
\kappa=\f{\hbar}{\delta l}.
\ee
Also $l_c$ being the radius of the cosmological horizon it doesn't make sense to talk about a momentum of norm smaller than $1/l_c$, so that the cosmological constant results in a resolution in momentum space:
\be
\delta p=\frac{\hbar}{l_c}=\hbar\sqrt{\Lambda}.
\ee
This effect can be compared to the quantization of speeds due to the presence of a cosmological constant occurring in the spin foam framework \cite{qspeed}.
\begin{equation}\nonumber
\begin{array}{c|c|c}
                           & \textrm{Space-time} & \textrm{Momenta} \\ \hline
\textrm{Curvature}         & \Lambda          & \delta l ^2 / \hbar^2 \\ \hline
\textrm{Resolution}        & \delta l         &  \sqrt{\Lambda}\,\hbar
\end{array}
\label{eq:table}
\end{equation}
We can try to formulate the argument in more mathematical terms. Let us consider a neighborhood $\oo$ of the point $x$
on the manifold $\mm$, of size (as measured with the flat Minkowski metric) very large compared to $\delta l$. Let us
consider a covering of $\oo$ with (open) balls $\oo_i$ of radius $\delta l$. Forgetting about the background manifold
and keeping only the collection of open sets $\{\oo_i\}$, we are left with a topological space we can embed in a flat
space-time $M$ which provides the effective description of the (curved) space-time for the observer with finite
resolution. Now following a geodesic on the initial curved space-time, we can trace the sequence of open sets $\oo_i$
it goes through. Such a geodesic, going in straight lines in the curved manifold $\mm$, would now be curved in the flat
manifold $M$. The observer defining momenta as generating motion, especially in a flat space-time where $\delta x$ is
simply proportional to the momentum $p$, he will describe the momentum space as curved. Of course, one could argue that
the curved motion is due to some acceleration, but we are considering free motion, or due to some curvature, but we are
placing ourselves in flat space-time, so that the only possible conclusion is that we have new physics which can be
effectively modeled by curved momenta. One could then say that it is possible to generate curved objects by integrating
infinitesimal flat objects. However, due to the finite resolution (on the momentum space), we are now dealing with
integrated momenta which can no more be considered as infinitesimal: they are group elements (of $\SO(4,1)$) and not
anymore Lie algebra elements (of $\lalg{so}(4,1)$).

The key point is that we do not have access anymore to infinitesimal structures, both on the space-time manifold with
the resolution $\delta l$ or on the momentum space $\delta p=\sqrt{\Lambda}$. To have curved trajectories/geodesics
generated by {\it finite size} momenta requires that they belong to a curved manifold. Mathematically, using the
resolutions $\delta l, \delta p$, there is actually a duality between the curved space-time structure $(\mm,T_x\mm)$
and the effective flat space-time $(M,\tt_xM)$ with curved \lq\lq tangent space". More precisely, defining the
exponential (bijective) map $f$ on a (open) neighborhood of $x$, taken as the fixed origin relative to which the
construction is performed, sending vectors $p\in T_x\mm$ to points in $\mm$, we can use its inverse to send \lq\lq
momentum vectors" in $\tt_xM$ to points of the flat space-time $M$. Indeed, $M$ and $T_x\mm$ are isomorphic up to
scales, as well as $\mm$ and $\tt_xM$: $(\delta l/\delta p)\times f^{-1}(\cdot\times\delta l/\delta p)$ maps vectors of
(a neighborhood of) $\tt_xM$ onto points of $M$. More on the map $f_\kappa$ will be said in section \Ref{sec:coo}.

\begin{figure}[h]
\label{fig1}$
\begin{array}{c}
 \includegraphics[width=7cm]{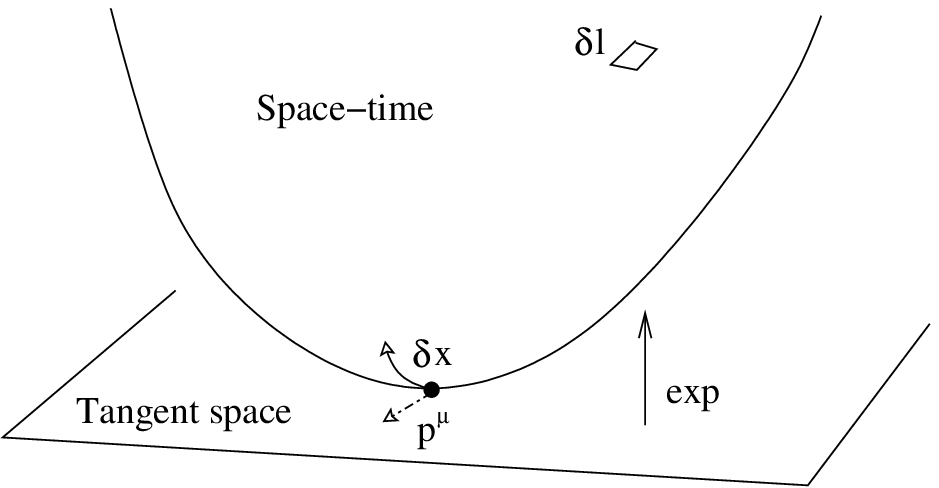}\end{array}{\bf \longrightarrow} \begin{array}{c}  \includegraphics[width=7cm]{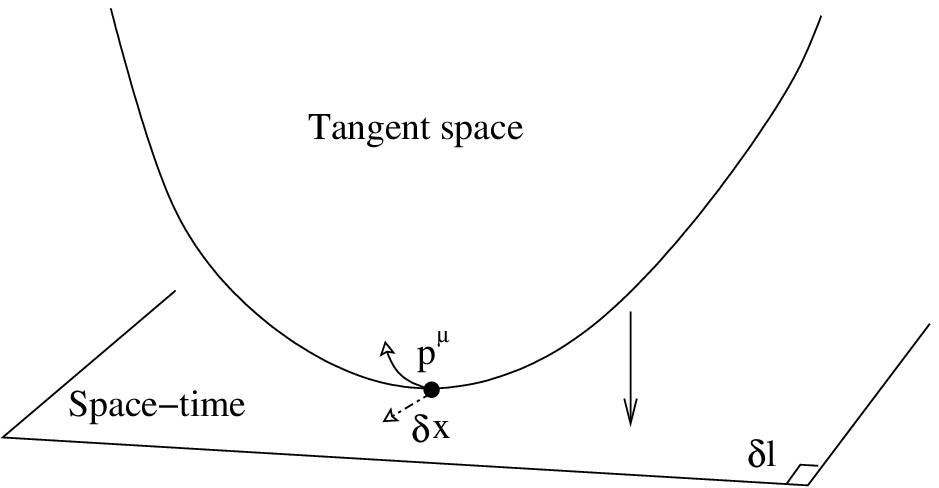}
 \end{array}$
\caption{Because we have a finite resolution, we can describe in the same way  a curved space-time (with flat tangent
space) and a flat non-commutative space-time (with dual curved momentum space).}
\end{figure}

\medskip

So we started from a coordinate space of constant curvature $\Lambda$ and resolution $\delta l$ with a flat tangent
space, from which we obtained a flat coordinate space with a momentum space of curvature $\frac{1}{\delta l}$ and
resolution $\sqrt{\Lambda} \hbar$.

Let us now consider the different limits, in order to check if everything is consistent. If the we have $\delta
l\rightarrow0$, this means that the observer is observing with a perfect resolution, the momentum space becomes then of zero curvature, that is flat. One recovers the Special Relativity case. If we have a renormalization group picture in
mind,  $\delta l\neq0$ means  we have integrated out the quantum gravitational effects with characteristic length scale shorter than $\delta l$, with their presence of now encoded in the curved momentum space.

When $\Lambda$ goes to 0, the resolution $\delta p$ goes also to 0 and can be neglected, so that momentum space becomes
a smooth manifold; let us stress that this is, from our perspective, an approximation that can be relaxed, since the
all construction is based on the fact that we have a non-zero, albeit small, cosmological constant; also, one can
imagine the observer to achieve the maximal resolution allowed by quantum gravity, that is $\delta l=l_P$; in this way,
all that is left is: a flat Minkowski space-time $M$ with a curved momentum space $T_xM$ of constant curvature
$1/\delta l=1/l_P$. Locally, the momenta space has of course a trivial topology, and being of constant curvature, it is
then isomorphic to de Sitter space (we are assuming a positive cosmological constant). This is exactly the structure of
the non-commutative space-time of deformed special relativity: from the de Sitter structure of the momentum space, one
can derive the commutators between the Lorentz generators and the momenta (translations) and also the commutators of
space-time coordinates \cite{snyder, jerzy}. One result is that the coordinates are then non-commutative. Then, at the
algebraic level, there is an ambigu\"\i ty in what one should call the space-time coordinates for we can change basis
without mathematical inconsistency i.e there is no preferred choice of coordinates (more precisely, we don't yet have a
definite physical motivation to choose one). We shall come back to this issue in the next section. The natural choice
is the Snyder's basis: \be [X_\mu,X_\nu]=-\f{i \hbar^2}{\kappa^2}J_{\mu\nu}=-i(\delta l)^2J_{\mu\nu}. \ee The (square)
length Casimir is expressed as $L^2=X_0^2-X_iX_i$. The resulting uncertainty relations read as:
$$
\delta X_\mu \delta X_\nu=(\delta l)^2 \la J_{\mu\nu}\ra,
$$
so that $\delta X\sim \delta l$, which is consistent with the starting point that one has a finite resolution $\delta
l$ on measurements on the space-time coordinates.

Finally, to sum up, we had an observer in a space-time of constant curvature $\Lambda$ working with a finite resolution
$\delta l$. The physics in such framework can be consistently described effectively by a flat non-commutative
space-time with a curved momentum space.

\medskip
 Let us comment on the origin of the (small) curvature $\Lambda$. $\Lambda$ seen as the energy density of
the vacuum can be considered as the most basic quantum gravity effect: quantum fluctuations of the metric/curvature
would locally assign a non-zero expectation value (though a priori small) to the scalar curvature. From this point of
view, looking at quantum gravity around the flat Minkowski background, quantum gravity effects would appear to curve
slightly space-time so that an observer would describe it as a non-commutative flat Minkowski space-time. Then the
uncertainty in measurements of space-time coordinates would be due to quantum fluctuations of the background metric,
whose precise values the observer can not have access to.
This, together with the fact that the resolution
$\delta l$, which is at the root of our argument, is bounded from below by
the Planck length and therefore cannot be neglected even in principle
because of quantum gravity considerations,  presents {\bf DSR as an
effective model for quantum gravity}, in the nearly flat limit.
Later, we will discuss how to extend DSR to arbitrary background curvature so that it would describe
effective physics over a space-time with arbitrary fluctuations of the curvature (and not only of its scalar
component).

The framework of DSR arise when this cosmological constant, which is
the physical origin, together with the finite resolution in
configuration space, of the curvature in momentum space, can be
approximately neglected  as regards to the finite resolution it would
imply in momentum space, thus allowing for a description of the latter
as a smooth manifold. Therefore another extension we are naturally led
to present is to take into account the finite $\Lambda$ case, by keeping track of the finite resolution on the momentum space. This will lead us to a notion of {\it Doubly Deformed Special Relativity}, where the momentum operators will also become non-commutative which will lead to a $\sqrt{\Lambda}\hbar$ uncertainty in measurements of momenta.

\medskip

Let us remark that our framework seems very close to the construction by Kempf \cite{sampling} using the {\it sampling theorem}. Indeed a
continuum is equivalent to some discrete information, when one has a cut-off, i.e. a finite resolution, the classical
space-time can be effectively described by a discrete amount of degrees of freedom. Interestingly, it is also the
point of view of the Seiberg-Witten map, which shows the equivalence of a quantum field theory on a non-commutative
space-time with one on a classical commutative manifold, when expanding the observables in powers of the deformation
parameter. We take exactly the same point of view: assuming a finite resolution, we  map the physics on a classical
space-time to the physics on a non-commutative space-time, by locally mapping the (discrete) space-time to its momentum
space (and reciprocally). Moreover, we identify the resolution (of the sampling theorem approach) with the quantum
deformation parameter. It thus seems very interesting to study the exact interplay between these three notions: DSR,
Sampling theorem and Seiberg-Witten map. It will certainly be enlightening in order to understand how to formulate a generic
effective and operational theory of quantum gravity. Indeed a truly mathematically rigorous version of our argument would be to prove that some quantum correlation functions or observables on the De Sitter space-time would coincide with some quantities computed in the DSR framework (at least at first order in $\hbar$ or $\delta l$). We leave this for further investigations.

\section{Addressing the issues of DSR}

We are proposing a new approach to deal with DSR, and this should give a new point of view to tackle some of the
problems encountered in DSR. There are, in our opinion, two main problems in DSR. The first one concerns the multitude of possible
deformations presented in the first section, related to different
choices of coordinates in momentum space: are they all physically equivalent or there is one pinpointed  by physics?
The second problem is the so called \lq\lq soccer ball problem". The space of momenta is deformed so that the addition of
momenta cannot give a momentum or energy greater than the Planck energy. However, macroscopic objects, like a soccer
ball, have an energy much greater than the Planck one, so it seems in complete contradiction with the initial hypothesis
of  the sum bounded by the Planck energy.

We shall see that the geometrical approach taken here allows to solve the first problem and gives new
insights on the second.

\subsection{Coordinates ambiguity or is only one deformation
  physical?}
\label{sec:coo}
From the algebraic approach, we have many possible deformations, as indicated in the equations (\ref{deformation1}),
(\ref{deformation2}), and one seems to be forced to select one as the
\lq\lq correct'' one. On the other hand when considering the Snyder's approach, that is the geometric approach, all the
deformations are related to some coordinates systems (possibly
singular), and from this geometric perspective all the  coordinates systems are equivalent to each other. This would imply then
that  there is not only one physical deformation, but many, physically
equivalent to each other. These two point of view (algebraic=
only one deformation is physical, geometric= all deformations are
equivalent) are both commonly heard in the community. Our
approach shows that in fact  there is a bit of truth in each of them,
and how to reconcile them.

We have introduced the map $f_{\kappa}$ which sends the flat tangent space $T_x\mm$ to the De Sitter Space $\mm$. The
map $f_\kappa$ that we are using is the exponential map between the subspace of the $\lalg{so}(4,1)$ lie algebra
generated by the translations ($J_{\mu 4}$) and the De Sitter space seen as the homogenous space $\SO(4,1)/\SO(3,1)$
(which is the exponential map between the Lie algebra $\lalg{so}(4,1)$ and the group manifold $\SO(4,1)$ quotiented by
the action of $\SO(3,1)$). It can be equivalently defined as the inverse of the map going from any coordinate system in
a neighborhood of any point in the manifold to the "normal coordinate system" in which the Christoffel symbols vanish,
so that space-time looks flat, in the particular case of De Sitter geometry. It has also the interpretation of an
exponential map. As the map $f_{\kappa}$ depends on $\kappa$, it encodes uniquely the deformation on which the DSR
approach is based, and does not depend or refer to any specific choice of coordinates, being independent of any choice.
Indeed we can  change of coordinates systems in the tangent space $T_x\mm$ without altering the any of its geometrical
properties, and then the map $f_\kappa$ would make these correspond to different coordinate systems on $\mm$. Different
coordinates systems in the tangent space are all equivalent provided they are related by a non-singular transformation
(diffeomorphism). In this sense, all the coordinate systems are physically equivalent, but there is only one
deformation, encoded in the map $f_{\kappa}$.

This distinction between change of coordinate system and of deformation can be understood as the difference between a passive and an active coordinate change: a passive transformation would be to describe the same physics using different observables, while an active transformation would assign the new different observables to the same original measurement process. Here DSR is a deformed special relativity and thus only invariant under passive coordinate transformations. We speculate that a deformed general relativity, which should correspond to a quantum theory of gravitation, should be invariant under active coordinate changes (i.e changes of the deformation). It would thus satisfy a generalized equivalence principle, which would be stronger than the classical one since it should take into account the quantum feature that an observer is not only defined by its position and motion but we also need to specify a set of measurements (a basis of the quantum algebra of observables).

Note that this is very close to what happens in going from Galilean
kinematics to Special Relativity \cite{SR}. Indeed in this case, we want to deform the space
of speeds (which is isomorphic to the space of momenta). Initially this space is isomorphic to $\R^3$, and to deform
it, so that there is a maximal speed,  we have to deform it as the hyperboloid $SO(3,1)/SO(3)$. The map
$g_c:\R^3\rightarrow SO(3,1)/SO(3)$ encodes uniquely the deformation\footnote{The speed of light $c$ is now the deformation
parameter or the curvature of the hyperboloid}, but there are still
many coordinate systems one can choose on the hyperboloid, leading to
different expressions for the quantities of interest. In particular,
it is shown in \cite{SR} that there are the Snyder-like coordinates or the so called Special-Relativity coordinates. These are
all physically equivalent once the deformation $g_c$ is known.

\subsection{Addition of momenta or the soccer ball problem}

In the DSR framework, the addition of momenta, and the notion of composite systems,
is not well understood: the usual linearity is {\it a priori} lost, and the sum of two momenta is constrained by
the formalism to be always bounded by the Planck energy $E_P$,
in clear contradiction with everyday experience of composite objects
made out of large numbers of elementary ones. More precisely, by
construction the DSR algebra, with its special co-product structure,
would not let the sum of two momenta or masses or energies, depending
on the chosen basis of DSR  exceed the Planck
scale.
And it finally seems that the DSR algebra only describes one-particle states and that we have to look beyond it
in order to describe multi-particle states. One way out was proposed by Magueijo-Smolin \cite{leejoao}, but still faces puzzling issues, most notably the non-associativity of the resulting definition of momenta addition.
We believe that the effective theory point of view helps to understand the macroscopic behavior of DSR theories.

\medskip

Let us start by reviewing the basic motivation for writing a DSR-like theory when considering a flat space-time theory taking into account some quantum gravity effects. A first motivation is to take into account the universality of the Planck length $l_P$: we would like a minimal (space-time) length or a notion of a quantum of length/distance.
In a flat space-time, through the Fourier transform, a quantum of
length leads to a characteristic scale in momentum space. The
corresponding Planck mass $m_P$ gives us a bound on mass or energy or
momentum depending of the particular precise notion of minimal
length. This resembles the second motivation which is to impose a
maximal mass, or a maximal mass/energy density. Indeed, considering
some phenomena occurring in a (space) region of length $L$, the corresponding Schwarzschild mass:
\be
M^{(L)}_{max}=2\f{GL}{c^2}
\ee
provides a bound of the mass/energy of the phenomena. Note that $M_{max}$ actually depends on the distance $L$, but is linear in $L$ so that it is not exactly an energy density. Now when $L=l_P$, $M_{max}$ goes to the Planck mass (more precisely $2m_P$) and we are lead to the usual DSR theories.

This second motivation shows us a way to go macroscopic masses much
larger than the Planck mass: we need to introduce another length scale
$L$ -the macroscopic scale- and renormalize the theory in order to
deal with macroscopic objects.  More precisely, from the effective
theory point of view, we have the resolution $\delta l$ of the
observer and the length scale $L$ of the observed phenomena. Standard
DSR theories describe the case $\delta l \sim L \sim l_P$. Then we
would like to renormalize the theory to go from the Planck scale to
the macroscopic scale. This is the issue of the (semi-)classical
limit: we would like to consider $L\gg l_P$, or more exactly $L/\delta
l \gg 1$. There is two equivalent ways to visualize the
situation. Either we keep $\delta l$ fixed and we take the large $L$
limit: this is the \lq\lq coarse graining" point of view, where you have a
fixed microscopic structure for the system whose large scale structure you study . Or
you choose a macroscopic scale $L$ and you take $\delta l$ very small:
this is exactly the framework of the \lq\lq classical limit". We here use this latter point of view: when $\delta l$ decreases and becomes very small, the deformation parameter $\kappa\sim 1/\delta l$ increases and grows very large. $\kappa$ being the characteristic mass  and bound on mass/energy/momentum, we are increasing the bound and sending it to $\infty$ in the limit $\delta l/L\arr 0$. In the earlier point of view when $\delta l$ is fixed and $L$ grows, we understand that $\kappa$ will increase (linearly) with $L$, just the same way that $M_{max}$ grows linearly with $L$. We tentatively write:
$$
\kappa\sim \f{\hbar}{l_P}\f{L}{\delta l}.
$$

More precisely, starting off with a system with
length scale $L$, one should cut it up in (space-time) regions of size
$l_P$, each governed by DSR and eventually interacting with each
other. Taking many copies of the same original $\kappa$-Poincar\'e
algebra and extracting from it the \lq\lq coarse-grained" deformed
Poincar\'e algebra, we should find a renormalized larger $\kappa$
parameter. This is actually shown in more details in
\cite{lorentz}. From the geometrical point of view, for two particles,
we start with two copies of the DSR algebra with two De Sitter
momentum spaces, i.e two one-particle states, and we build the
corresponding one two-particle state. If the original masses/energies
are bounded by $\kappa=m_P$ (in the center of mass frame) then the
total mass/energy will be bounded by $2m_P=2\kappa$. Then physically,
intuitively, we want  to take the De Sitter momenta for each particle
and map them to the \lq\lq true" ideal flat momenta using the map $f_\kappa$, sum them in the flat momentum space, and then map it back to a De Sitter momentum this time using the coarse grained map $f_{2\kappa}$:
$$
p_{tot}=f_{(2\kappa)}(f^{-1}_{(\kappa)}p_1+f^{-1}_{(\kappa)}p_2).
$$
This is actually the Magueijo-Smolin proposal \cite{leejoao}, except than they place themselves in the algebraic
framework using unitary operators while we use the map between the tangent space and the curved manifold and exploit
the scale ambiguity arising from the fact that the space-time scale and the momentum scale are {\it a priori}
independent. The details of the coarse graining at the level of the algebras and the precise relation\footnotemark
between the original momenta and the total momentum can be found in \cite{lorentz}.

\footnotetext{The coarse-grained deformation parameter $\kappa'$, here
  set to $2\kappa$, is actually to be computed more precisely and
  depends {\it a priori} on the details of the relative motion (and
  interaction) of the two \lq\lq particles".}

Then, since this \lq\lq total momentum addition" is different from the momentum addition corresponding to the co-product of the $\kappa$-Poincar\'e algebra, it is natural to wonder what is the physical meaning of the Hopf algebra sum rule.
It corresponds to change of reference frames: if the mass/energy is to be bounded in one given reference frame, then it will be bounded in all references frames and the law ruling the behavior of the momenta (symmetry algebra) should be the same in all reference frames  as stated by the relativity principle. This point of view is to be compared to the  Special Relativity case as presented in \cite{SR}. First we have the speed addition which is unbounded so we deform
the space of speeds in the hyperboloid so that the sum stays smaller than the speed if light. We can add as many speeds
we wish (that would be the analogous to add many momenta to get a momentum of energy higher than $E_P$), the sum will
never be higher than the speed of light. In fact what is bounded is the galilean speed $\overrightarrow{v}$, and the
relativistic speed $\gamma \overrightarrow{v}$, which defines the momentum, will not be bounded. In fact one can understand that the galilean speeds
(associated to the Galilean symmetry) are bounded with respect to the new symmetry (Lorentz), but the relativistic
speeds (associated  to the Lorentz symmetry) are not. This prompts the intuitive guess: there must be for DSR a new
kind of momentum associated to the new symmetry ($\SO(4,1)$), which should be the right quantity to add in order  in
order to get the right limit. Then the sum rule for momenta -thinking of scattering- is {\bf not} given by the speed sum rule: one deals with change of reference frames, the other with scattering and total momentum.
In fact this renormalization methods provides is an intuitive way to implement
this new momenta and study its behavior. This is indeed showed in a forthcoming paper \cite{lorentz}.

As we have seen,  the framework of effective theory is very efficient in understanding the physical content of DSR theories, and the notion of multi-particle states.

\section{Towards an effective Quantum Gravity model from DSR}

Up to now we have been considering a space with a constant curvature, and we have showed how a curved space-time
together with a minimal length can be interpreted as a non-commutative
space-time, leading to the DSR framework. The idea
would be now to interpret the DSR as reproducing/representing the quantum fluctuations of space-time and thus to take
into account small bumps of curved space-time. These bumps need not be only in the scalar curvature, so that we need to
generalize our effective theory method to describe physics on any a priori arbitrary curvature: we would like a flat
space-time with a non-commutative structure taking into account the effects of a small arbitrary curvature. In this
sense we would develop an effective theory of quantum gravity in the
small curvature regime. As said previously, the first step in
this direction is to extend the DSR framework to space-time with varying curvature and not only of constant curvature.
This would naturally imply having a varying non-commutativity parameter $\kappa$ and even generalize it from a scalar
to a tensor.

To this purpose, we follow the same method as in the case of a space-time of constant curvature. Assuming that the
curvature is small at the resolution of the observer, the observer can once again consider the space-time to be
effectively flat. However small displacements around the observer,
located at $x$, say, take place in the true
space-time ${\cal M}$. Then, because of the finite resolution, one can not consider infinitesimal displacements, so
that the effective momentum space becomes curved as the original space-time is.

From the structure of curved momentum space, one can in principle reconstruct the commutation relations of space-time
coordinates, which should be expressed in terms of the metric $g$ in
momentum space and $\kappa=1/\delta l$. We leave their exact expressions for
future investigation, giving just some suggestions below.

Let us recall now some facts about scale in GR. The GR equations of motions are globally scale invariant, and moreover,
pure gravity is also scale-free. These are two different notions as
globally scale invariance does not imply scale-freeness.
When introducing matter, we are introducing a intrinsic scale (e.g. rest mass), but we can still have globally scale
invariance.  Local scale invariance is a much stronger constraint (requiring it makes the theory conformally
invariant). This local scale is given by the conformal factor.
As we have associated the scale to an observer, the conformal factor renormalises the reference unit of the observer,
or equivalently, represents how the scale of the geometry changes according to the chosen unit.
We insist on the fact that the observer keeps her own unit/ruler fixed, and it is the rest around her that changes and gets scaled.

In fact we can see that by mixing the  dynamical point of view and the kinematical point of view we have an effective
notion of this minimal resolution, which will encompass $\delta l$ and the conformal factor.

Dynamics is a necessary step to consider as indeed if  we have generalized DSR to describe any curved space-time,
 general relativity, and thus quantum gravity, also describe the dynamics of space-time: one would like to
describe how the effective non-commutative structures seen by different observers are related, or equivalently how the
non-commutative description given by an observer is going to evolve. Note that one interesting feature of this
generalized DSR is that the non-commutative space-time structure seen by one observer (at a certain point $x$) is going
to differ from the one seen by another observer (at another point $y$). This hints towards the idea of a relative
non-commutative geometry \cite{relatncg}, in which the non-commutative structure depends on the observer and also the
resolution which he is using to  probe space-time.

Let us look more closely to the splitting conformal factor-conformal metric $g=e^{\phi}\what{g}$,  with
$det(\what{g})=-1$.  The action for General Relativity reads then as\footnotemark:
$$
S_{GR}=\f{1}{(\delta l)^2}\int d^4x \,\sqrt{det(g)}R= \int d^4x\,\sqrt{det(\what{g})} \,
\f{e^{3\phi}}{(\delta l)^2}\,\times\left(\hat
R - 3\hat g ^{ab}\nabla_a \nabla_b\phi- \f{3}{2} \hat g^{ab}\nabla_a\phi \nabla_b\phi\right).
$$
\footnotetext{The action for General Relativity has the factor $c^3/G$
  in front of it. In a path integral formulation, the amplitudes are
  computed from a dimensionless exponent, thus with the factor
  $c^3/\hbar G=1/l_P^2$ in front of the action. Note that these factors will not modify the equations of motion but will affect the path integral, and will especially change under renormalisation. More precisely, at the effective level, we renormalise the Planck length factor to the resolution $\delta l$ associated to the observer.}
From the term on the right-hand side, we can see that the minimal
resolution can be redefined by encompassing the conformal factor,
leading to an effective resolution $\delta l_{eff}$. From this point of view, the observer sees  the change of scale of the space-time around her as given by the scalar field $\phi$, whereas the conformal metric is giving the geometry. As a consequence one can conjecture  that the new commutation relation between the space-time coordinates will be of the form
\begin{equation}
[X_{\mu}, X_{\nu}]=-i \kappa(\delta l, \phi)j_{\mu\nu}(\what{g}_{\mu\nu}),
\end{equation}
where the metric $g$ can now be seen, according to the argument we
presented above, as the non-trivial metric induced in momentum space
by the presence of a finite resolution.
In this way, the dynamics of $\phi$ defines the dynamics of the deformation parameter $\kappa$.
Let us point out that changes of $\kappa$ in DSR is achieved by applying a conformal transformation on the
whole Poincar\'e algebra (see for example \cite{leejoao}).
The dynamics of the (conformal) metric $\what{g}$ leads to a dynamical $j_{\mu\nu}$, which is
the generalization of the  Lorentz generators $J_{\mu\nu}$. More precisely, identifying the the coordinate
operators $X_{\mu}$ as the generators of the translations on the curved momentum space (Lie derivatives), we can express their commutators using the Riemann tensor\footnotemark associated to the metric $\what{g}$ on the momentum space:
$$
j_{\mu\nu}=R_{\mu\nu}{}^{\alpha\beta}[\what{g}]\,J_{\alpha\beta}.
$$
Note that this reduces to the Snyder form of DSR in the case of De
Sitter geometry as it should.
An extra-subtlety is that the observer fixes the scale (she assumed her unit $\delta l$ always unchanged and measures everything in terms of it), and the conformal factor should only affect the scalar $\kappa$ and not the generators $j_{\mu\nu}$ governing the non-commutativity of the algebra. Therefore, we should use the Weyl tensor $W$, which is the traceless part of the Riemann tensor and invariant (or more precisely covariant) under conformal transformation, so that we have at the end of the day:
\be
[X_{\mu}, X_{\nu}]=-i \kappa(\delta l, \phi) W_{\mu\nu}{}^{\alpha\beta}[\what{g}]\,J_{\alpha\beta}.
\ee
Then, from the point of view of the observer, the conformal factor $\phi$ should manifest itself simply as an extra dynamical scalar field.

\footnotetext{The Riemann tensor $R_{\mu\nu\alpha\beta}$, considered as a matrix $(R_{\mu\nu})_{\alpha\beta}$, is defined as the commutator of the parallel transport of a vector $v_\alpha$ along the $\mu$ and $\nu$ directions:
$$
\left([{\cal L}_{x_\mu},{\cal L}_{x_\nu}]\cdot v\right)_\alpha=R_{\mu\nu\alpha\beta} v^\beta.
$$}

Note that this structure is very close to the unimodular gravity introduced by Anderson and Finkelstein
\cite{unimodular}. Also, the idea of a generic curved geometry in
momentum space as an effective description of (some) quantum gravity
phenomena was suggested by Moffat \cite{moffatmom}. We hope to be possible to obtain an effective
theory of quantum gravity (effective meaning \lq\lq as seen by an observer") on these bases.

\medskip

We should point out that neither the Riemann tensor, nor the Weyl
tensor, are invariant under diffeomorphisms, they change under simple
coordinate changes. Then they really depend on the observer, whose
description of physics we are considering: an accelerating observer
will see a different Riemann tensor and would derive different
commutators between the coordinates. However, such modifications still
provide isomorphic (quantum) algebras. Everything resides in what one
calls \lq\lq coordinates" and considers as physical quantities that can be effectively measured.

Finally, let us remind that we are here taking the point of view that
the non-trivial metric and curvature tensor used here already include
the quantum fluctuations of the metric (around some given classical
geometrical background), and are assumed to be of small magnitude, and
that the non-commutativity of the coordinates results from the fact
that the observer does not have direct access to these metric and
curvature tensor (being hidden from her by her finite resolution
available) and still attempts to describe physics as she was in a flat space-time.

\medskip

From the Magueijo-Smolin algebraic point of view \cite{leejoao}, the generalized DSR theories should be obtained by
applying a unitary transformation to the Poincar\'e generators, $J,p\arr UJU^{-1},UpU^{-1}$, and we would be studying
the dynamics of this unitary transformation $U$. Interpreting $U$ geometrically, it corresponds to the map $f$ between
the flat momentum space and the curved momentum space, or equivalently between the curved space-time and its tangent
space. Thus $f$ describes the metric of the space-time and it is natural to make it dynamical in the context of an
effective quantum gravity model.

As a final remark, let us comment on the Poincar\'e invariance in this setting. Poincar\'e invariance is a fundamental
feature of the DSR approach. At the effective level, we have a flat space-time $M$ with a curved momentum space $T_xM$.
The action of the Poincar\'e group on $M$ is obtained by applying the map $f$ from $T_x{\cal M}$ to ${\cal M}$
(defined in a neighborhood of $x$). This maps Poincar\'e symmetry of $T_x{\cal M}$ to (local) diffeomorphisms on ${\cal M}$. So that the effective Poincar\'e symmetry of $M$ (identified to $T_x{\cal M}$) is given by modifying the
generators $J,P$ to $fJf^{-1},fPf^{-1}$. This way, translations becomes non-commutative and the action of the Lorentz
group becomes non-linear.
However, General Relativity, taken in its first order formulation in terms of tetrad and spin-connection, doesn't seem
to be invariant under Poincar\'e but only under the Lorentz group. So how can we hope to use these Poincar\'e invariant
construction to build a quantum theory of gravity? Actually, there is locally an action of the Poincar\'e group: the
Poincar\'e group naturally acts on the tangent space and can be mapped to diffeomorphisms on the original
manifold in a neighborhood of any given point. The resulting action of the Poincar\'e group will obviously be non-linear,
which is consistent with our proposition. Horowitz has shown indeed that on-shell, the gravity action is invariant
under infinitesimal Poincar\'e transformation \cite{horowitz93}. In
other words, one can consider DSR as a {\emph local} effective
description of quantum gravity, with the deformed Poincar\'e
used in it representing the non-linear action of local diffeomorphisms.

\section{Taking the effective viewpoint further: a Doubly Deformed Special Relativity}
\label{sec:DDSR}

The way we argued DSR arises from a Quantum Gravity theory in an effective description of it was based on the following
basic points: 1) Quantum Gravity provides us with an irreducible minimal scale of resolution $\delta l$ for
constructing our picture of space-time; 2) Quantum Gravity also produces a non-zero cosmological constant $\Lambda$
considered as a first order effect of quantum fluctuations of the gravitational field, or as a renormalization of any
bare cosmological parameter included in the action; 3) this cosmological constant corresponds to a maximal length scale
which is however much larger than that at which our experiments take place and even more than the minimal quantum
gravity scale, therefore an effective description in terms of a flat space-time makes sense in spite of the presence of
$\Lambda$; 4) however, the presence of a minimal resolution in configuration space produces a curvature in momentum
space proportional to it, and a geometry which mimics the one in configuration space, i.e. De Sitter; as a consequence,
the space-time coordinates, identified with generators of translations on momentum space, become non-commutative, so
space-time becomes fuzzy, and physics is described by a Lorentz invariant non-commutative geometry on which a deformed
Poincar\'e algebra acts, with a deformed action of boosts on space-time coordinates, i.e. by a DSR theory.

There is a further ingredient of this picture that we mentioned above, but didn't stress enough: the presence of a
cosmological constant $\Lambda$, by the same line of argument we followed analyzing the effect of $\delta l$, induces a
minimal resolution in momentum space; we neglected this second sort of minimal resolution on the basis of the smallness
of $\Lambda$ as compared with the other quantities appearing in the model; as a consequence we could describe momentum
space as a smooth (curved) manifold, with a De Sitter group of symmetries, giving a non-commutative space-time.

However, this is a further approximation that, although justified, may be dropped. In the light of the points recalled
above, a more precise effective description of flat quantum gravity must take into account both the two scales induced
by the full quantum gravity theory; the way to do it is to drop both sorts of smoothness one has in a classical
description of space-time geometry: of space-time itself and of momentum space, the first made impossible by the
minimal resolution $\delta l$, the second by the minimal momentum resolution $\sqrt{\Lambda}$.

The question is how to do it. If the above heuristic arguments give a motivation for assuming a certain description,
for the non-commutativity of space-time and the deformation of flat space symmetries, and for the properties of
momentum space, they do not give us the details of this description when the smoothness of momentum space is not
assumed anymore. The only thing we can rely on, at this level, is just the symmetry in the description of configuration
space and of momentum space arising from that line of argument and well summarized by the table  \Ref{eq:table}, and
then try to build up a similarly symmetric double non-commutative structure for both spaces, where the two scales
$\delta l$ and $\sqrt{\Lambda}$ play a similar formal role.

The way the minimal resolution in space-time enters in the final non-commutative description of it is in the
commutation relation for space-time coordinate operators $X_\mu$, which in the Snyder's basis takes the form:
\be
\left[ X_\mu, X_\nu\right]\,=\,-\,i\,(\delta
l)^2\,J_{\mu\nu}. \label{eqn:snyder}
\ee
This, together with the standard commutators between Lorentz algebra generators,

\be
\left[ J_{\mu\nu} , J_{\rho\sigma}\right]\,=\,\eta_{\nu\rho}\,J_{\mu\sigma}\,-\,\eta_{\mu\rho}\,J_{\nu\sigma}\,-\,\eta_{\sigma\mu}\,J_{\rho\nu}\,+\,\eta_{\sigma\nu}\,J_{\rho\mu}
\ee

implies a deformed action of the Lorentz group on space-time, with the usual action of the generators:

\be
\left[ J_{\mu\nu} , X_\rho\right]\,=\,i\, \left( \eta_{\mu\rho}\,X_\nu\,-\,\eta_{\nu\rho}\,X_\mu\right),
\ee

as we are requiring that the 10 operators form a De Sitter
Lie algebra (of symmetries of momentum space).

How can we obtain a purely algebraic picture of a non-commutative space-time {\it and} a non-commutative momentum
space? The presence of a minimal resolution in momentum space coming from the cosmological constant, and appearing just
as the minimal resolution $\delta l$ appears in the space-time sector, strongly suggests a commutation relation among
momentum coordinates of the form:

\be \left[ P_\mu, P_\nu\right]\,=\,-\,i\,\hbar^2\,\Lambda\,J_{\mu\nu}, \ee  i.e. fully analogous to the space-time
sector of the algebra, to be considered together with the above; here we are also assuming a Snyder choice of
coordinates for momentum space, corresponding to the choice of coordinates in the space-time sector we have made in
\Ref{eqn:snyder}.

The cosmological constant can be seen thus as introducing a new
distance scale $(\Lambda)^{-1/2}$ or, equivalently, a new momentum
scale $\Pi = \sqrt{\Lambda} \hbar$ (so that $\left[ P_\mu,
  P_\nu\right]= - i \Pi^2 J_{\mu\nu}$).

Again on the basis of symmetry between configuration and momentum
space, we can require also the 10 operators given by the 4 momenta and
the six Lorentz generators to form a De Sitter algebra,

\be
\left[ J_{\mu\nu} , P_\rho\right]\,=\,i\, \left( \eta_{\mu\rho}\,P_\nu\,-\,\eta_{\nu\rho}\,P_\mu\right),
\ee

thus imposing the same form of action for the Lorentz subalgebra on
momentum space, implying again a deformed action of the group on
momentum coordinates. This is of course consistent with the Snyder choice of
coordinates in momentum space.

Under these assumptions, and using the above commutation relations, it is a straightforward although tedious exercise
to derive the remaining commutation relations between the space-time and momentum coordinates; the most general form
for these that is allowed by covariance is:

\be \left[ X_\mu , P_\nu \right]\,=\,
A\,\eta_{\mu\nu}\,+\,B\,J_{\mu\nu}\,+\,C\,X_\mu P_\nu\,+\,D\,P_\mu
X_\nu\,+\,E\,P_\mu P_\nu\,+\,F\,X_\mu X_\nu    ; \ee  the exact form of the coefficients in this expression can then be found by using the various Jacobi
identities. The result is:

\be \left[ X_\mu , P_\nu \right]\,=\, \hbar \eta_{\mu\nu}\,-i\,\delta
l\,\sqrt{\Lambda}\hbar\,J_{\mu\nu}\,-\,\sqrt{\Lambda}\,\delta
l\,\left( X_\mu \,P_\nu\,+\,P_\mu\, X_\nu\right)\,-\frac{(\delta
  l)^2}{\hbar} P_\mu P_\nu\,-\,\Lambda \hbar X_\mu X_\nu \ee

Note that it goes to the usual expression for the same
commutator in the Poincar\'e algebra case for $\delta l \rightarrow 0$, and to the DSR case for $\Lambda \rightarrow
0$, as it should be the case.

Using dimensionless coordinates $p_\mu=\frac{P_\mu}{\Pi}$ and
$x_\mu=\frac{X_\mu}{\delta l}$, the full algebra takes the simplified
form:

\bes
\left[ x_\mu, x_\nu\right]\,=\,-\,i\,J_{\mu\nu} \hspace{3cm} \left[ p_\mu,
  p_\nu\right]= - \,i\, J_{\mu\nu} \\
\left[ J_{\mu\nu} , x_\rho\right]\,=\,i\, \left(
\eta_{\mu\rho}\,x_\nu\,-\,\eta_{\nu\rho}\,x_\mu\right) \hspace{2cm} \left[ J_{\mu\nu} , p_\rho\right]\,=\,i\, \left( \eta_{\mu\rho}\,p_\nu\,-\,\eta_{\nu\rho}\,p_\mu\right)\\
\left[ x_\mu , p_\nu \right]\,=\, \frac{\hbar}{\delta l \,\Pi} \eta_{\mu\nu}\,-i\,J_{\mu\nu}\,-\,\frac{\delta l \,\Pi}{\hbar}
\,\left( x_\mu \,p_\nu\,+\,p_\mu\, x_\nu\,+\,p_\mu p_\nu\,-\, x_\mu x_\nu\right)
\ees

where the symmetry between space-time and momentum space is even more apparent.

The end result is a Lie algebra with two scales, $\delta l$ and $\Lambda$ (or $\Pi$), and two modified De Sitter subalgebras, one
for the space-time sector and one for the momentum sector, each modified by the appearance of one of the two scales
above; both scales on the other hand appear in the cross commutator between the two sectors. In a sense the structure
above is that of two copies (configuration and momentum) of De Sitter space, where the latter is described in terms of
its symmetries only, thus in purely algebraic terms. Of course, the speed of light still plays the role of an invariant
scale as in ordinary special relativity and in DSR. The latter is
recovered, in the Snyder basis, if one takes the limit $\Lambda \rightarrow
0$, but not the limit $\delta l \rightarrow 0$.

This algebraic structure with three invariant scales is
currently being investigated by J. Kowalski-Glikman and L. Smolin
\cite{LeeJurekDDSR} (see also \cite{LeeTalk}) (and from whom we first learned about it), exactly
as a further deformation of DSR, and it is indeed of great interest on
its own.

We may name the resulting new algebraic model for the effective kinematics of
quantum gravity in a maximally symmetric, and almost flat, state as
``Doubly Deformed Special Relativity'' (DDSR). It is best seen as a (further, with respect to DSR) deformation of the phase space
of special relativity; in fact the latter has the structure of two
copies of the Poincar\'e algebra (position and momentum space) with the
relation of the momenta and coordinates being given by standard
duality, i.e. $\left[ X_\mu, P_\nu\right]=-i\hbar \eta_{\mu\nu}$; the
deformation of DSR amounts roughly to a deformation of the coordinate
space to a De Sitter algebra, leaving the momentum space commutators
unchanged, but with a consequent deformation of the duality relation;
DDSR re-establishes the symmetry between the two sector by making each
of them a De Sitter algebra and further deforming the duality
relation, as it is clear by the full DDSR algebra written above.

Let us also stress that it is very likely that the momentum sector of this DDSR algebra corresponds simply to a quantum
deformation of the De Sitter algebra in a classical looking basis, i.e. to a $SO_q(4,1)$ Hopf algebra, as it is quite
clear from the deformed action on space-time coordinates; the parameter $q$ should be identified with $\frac{\delta l\,
\Pi}{\hbar}$ and the DSR limit corresponds indeed to the simultaneous limit $\Lambda \rightarrow 0$ and $q \rightarrow
0$ with $\delta l$ (or $l_P$) kept fixed. We do not have a solid proof of this (work on this is in progress
\cite{LeeJurekDDSR}), but if obtained, it would  confirm the conjecture widely discussed in the literature (see
\cite{dsrcosmo}) that the symmetry of the perturbative vacuum in quantum gravity is indeed $SO_q(4,1)$ and that a
$\kappa$-Poincar\'e symmetry arises in a suitably taken limit of small cosmological constant, together with the DSR
effective description.

There are two main nice features of this new model, we think: first of all, as we said, it appears the natural
framework for an effective description of the kinematics of particles in an almost flat maximally symmetric quantum
state of gravity; in fact it follows directly from the first two of the starting points listed at the beginning of this
section, together with the assumption of maximal symmetry; second, and maybe most important, it may well have testable
consequences, just as the introduction of a length scale leading to DSR, as first remnant in the flat approximation of
quantum gravity, is the basic source of much quantum gravity phenomenology.

The general idea would simply be that the modifications coming from a full quantum gravity theory to our effective
theories in the weak gravity limit may be infrared as well as ultraviolet, i.e. affecting both the very low energy
scale as well as the ultra-high energy sector. These modification may be taken into account as a first step by a
modification of particle kinematics, which basically means a modification of the symmetry algebra realized in the
effective theory, instead that by different matter content of the same. As a second step, this would then be refined to
a modification of the effective dynamics of the gravitational field, i.e. of space-time geometry, already in the low
curvature limit. Many recently proposed explanations of cosmological or astrophysical puzzles, alternative to dark
matter or dark energy, can be seen, in our opinion, from the same perspective.

One example of modified gravitational dynamics, that can potentially
replace the need for dark matter and dark energy models, is Moffat's
non-symmetric formulations of gravity \cite{moffat1}, and it
would be very interesting to see whether the modified kinematical algebra
of the DDSR kind can provide the basis for such modified dynamics.

Another example could be the following: if one sees the scale introduced by the cosmological
constant as inducing a minimal acceleration scale $a_0 = \sqrt{\Lambda}c^2$ for
gravitational dynamics, as suggested also in \cite{LeeTalk}, then this scale is
of the same magnitude of that introduced in Modified Newtonian
Dynamics (MOND) models of galactic dynamics, again as an
alternative to dark matter \cite{MONDReview}.
Of course this is at present just a speculation, as there are many unclear
issues in relating this algebraic and special relativistic setting to the one of MOND, and there are many problems in
the same MOND approach to galactic dynamics (problems of compatibility
with General Relativity, mainly\footnotemark, see \cite{MONDReview,BekensteinMond}).

However, we think it is already a hint of the possible fruitfulness of studying this Doubly Deformed Special Relativity (DDSR)
in more details. Clearly all the motivations one may have for studying
DSR, i.e. the rich possible phenomenology of quantum
gravity it gives raise to, hold true for developing and analyzing more DDSR as well, as none of the DSR phenomena is
lost (for what we can see at least), e.g. modified dispersion relations, new thresholds in relativistic kinematics,
etc., and possibly new ones are to be considered, those arising from
the new scale introduced in DDSR.

\footnotetext{There are problems and difficulties involved in making
  DSR (and therefore DDSR) compatible with General Relativity as well,
or in other words in constructing a version of General Relativity in  which a minimal length scale is present and the
group of isometries of flat space is just as in DSR; we wouldn't go as far as suggesting that the two problems are
related since this suggestion would not be based on any solid argument, but nevertheless we find this last possibility
intriguing.}

In particular, interpreting the minimal resolution in momentum space as a sort of quantization of momenta (thus dropping, as
we have said, the description of momentum space as a smooth classical manifold), and thus of velocities, it seems to us that the
heuristic argument we are proposing furnishes a conceptually independent reason for the fact that the introduction of a
cosmological constant leads to a quantization of velocities in an effective description of quantum gravity, as it
happens in spin foam models \cite{qspeed}; in these models a minimal length scale also exists (at least in some cases),
so that we may see, using our argument,  the DDSR framework as a more complete effective description of spin foam
models with cosmological constant in the maximally symmetric case. The possible experimental signatures of this type of
effects mentioned in \cite{qspeed}, may therefore be referred also to a DDSR framework.

The fully symmetric way in which momentum space and space-time are treated in the DDSR algebra is also suggestively
reminiscent of the approach of \cite{frederick, caianiello, toller} to a modification of special relativity, leading to
the introduction of a new invariant acceleration scale, again with many possible experimental consequences. The
possible links between this two approaches deserve, we think, further attention.

These are at present not much more than speculations; anyhow, we stress again that the very fact that our argument,  however heuristic it may be, suggests and leads to
the investigation of a new type of effective quantum gravity models
shows in our opinion its fruitfulness.

\section*{Conclusion}

We have shown how to interpret a curved space with a minimal resolution as a non commutative space, by using the tools
coming from DSR theory. We saw first that it gives a new argument on why DSR can be considered as a flat limit of
quantum gravity. This interpretation of DSR coming from a curved space explains as well some of the problems of DSR.
The first one is to know if different deformations lead to the same physics or not, and their links with the different
coordinates systems. From the point of view we took, it is clear that there is only one deformation, and that all the
coordinate systems, given this deformation, are equivalent (which is expected from the geometric point of view). There
is a striking link between this and the notion of passive and active diffeomorphisms which prompted us to conjecture a
\lq\lq quantum equivalence principle".  The other problem concerns the addition of momenta: it is not clear how we can
get a classical limit (i.e. energy much larger than the Planck energy) from the addition of bounded momenta. We saw
that the non-linearity of the momenta addition is naturally explained from the geometric perspective: it encodes the
fact that a non-zero curvature is present. The introduction of the resolution scale gave also more physical content to
the formula representing the addition proposed by Magueijo and Smolin.

We then proposed a number of ideas  which deserve more attention and work. We list them here.

The first is the link between the DSR structure with the Sampling theorem (finite resolution) and the Seiberg-Witten
map (commutative space $\rightarrow$ non commutative space). If there is one it would imply a nice link between the
Sampling theorem and the SW map, which would be very interesting for a physical interpretation of the SW map. These
different links deserve to be clarified.

$$
\begin{array}{rcl}
 &  \textrm{Deformed Special Relativity}  &   \\
 ? \nearrow   \swarrow &  &  \searrow\nwarrow  ? \\
\textrm{Sampling theorem} & \stackrel{?}{\rightleftarrows} & \textrm{Seiberg-Witten map}
\end{array}$$

Usually one proposes a \lq\lq bottom-up" approach to describe quantum
gravity, i.e. consider the \lq\lq true" quantum gravity
theory and then make some approximations to get the low energy description. Here as an effective theory we started from
the simplest phenomenology of QG (a minimum length and a small but non-zero
cosmological constant) and then moved from there (\lq\lq top-down"): we proposed a generalization of the DSR theory to make an effective description of quantum gravity.
Indeed, we proposed to extend our approach to
space-time with non constant small curvature (curvature=quantum fluctuations), which would then generate a non constant
deformation parameter. From a scalar it would be then generalized to a tensor. This would be an effective description of the
kinematics of quantum gravity.  As quantum gravity is really about dynamics, one should really introduce in  this
structure some dynamics (in momentum space, see \cite{moffatmom}). We sketched the first step by splitting the metric into the conformal metric  and the
conformal factor, which is related to the scale of resolution. This splitting gives the first intuition of the theory
we want to have, however a lot has still to be done, especially to determine the correct formulation of the dynamics.
The deformation parameter  represents the quantum fluctuations of a (flat) space-time. This fits in a natural framework
as one knows that general relativity can be rephrased locally as a Poincar\'e gauge theory. The introduction of the
minimum scale implies therefore that we have a non-linear realization of this Poincar\'e symmetry, in accordance to the DSR philosophy.
In this sense, we should have a dynamical non-commutative
geometry. This can be compared for example with a particle in a
magnetic field. As shown by Jackiw \cite{J}, the magnetic field generates a non commutative structure for the particle and if we add the dynamics for the magnetic field then the non-commutativity becomes dynamical.
Then in the set-up of a dynamical DSR, possibly modeling quantum gravity,  it would be natural to introduce a
generalized equivalence principle stating that the physics seen from different observers, in different states
of motion and performing different measurements, should be the same even if the observers  would be describing it
using different effective models based on different deformations of the Poincar\'e algebra.
Such a quantum geometry extension of the equivalence principle should  be compared with the framework of Rainbow
gravity \cite{dsrgr} and with the already introduced concepts of a quantum equivalence principle.
\cite{qequiv}

Last but not least, we saw that our construction allows for a curvature and a minimal resolution, both in space-time
and momentum space. This construction is then close to the one introduced by Moffat \cite{moffat}. It is also similar
to the \lq\lq Triply Special Relativity" presently studied by
Kowalski-Glikman and Smolin \cite{LeeJurekDDSR}. Note that this structure is then
different than the DSR construction, in some sense it is a different
limit of Quantum Gravity, as mentioned in the introduction and
discussed in section \Ref{sec:DDSR}. We
conjectured that it could provide a new starting point for tackling
the Dark matter problem following the line of the MOND approach.

We leave all these different ideas  for further investigation.

\section*{Acknowledgements}

We thank Laurent Freidel for many discussion about DSR and about the action of diffeomorphisms and
Poincar\'e group in general relativity. We also thank Jurek
Kowalski-Glikman, Lee Smolin and Gianluca Mandanici for discussions
and useful comments.

\end{document}